\documentclass[12pt]{iopart}
\usepackage[bookmarks=false]{hyperref} 

\usepackage{iopams}
\usepackage{subfig}  
\usepackage{graphicx}
\usepackage{cite}
\begin{document}

\title{Photon cooling by dispersive atom-field coupling with atomic postselection}

\author{Felipe Oyarce$^1$ and Miguel Orszag$^{1,2}$}

\address{$^1$ Instituto de F\'{i}sica, Pontificia Universidad Cat\'{o}lica de Chile, Casilla 306, Santiago, Chile}
\address{$^2$ Centro de \'{O}ptica e Informaci\'{o}n Cu\'{a}ntica, Universidad Mayor, Camino la Pir\'{a}mide 5750, Huechuraba, Santiago, Chile} 
\ead{miguel.orszag@umayor.cl}
\vspace{10pt}
\begin{indented}
\item[]April 2019
\end{indented}

\begin{abstract}
We propose, in a Ramsey interferometer, to cool the cavity field to its ground state, starting from a thermal distribution by a dispersive atom-field coupling followed by an atomic postselection. We also analyze the effect of the cavity and atomic losses. The proposed experiment can be realized with realistic parameters with high fidelity.
\end{abstract}
In the realm of cavity QED, a variety of experiments developed for generating nonclassical states of light assume that the cavity field is initially in its vacuum state \cite{nogues, colloquium, krause, QND90, QND91, QND92, domokos, varcoe, dotsenko, voguel, moussa, serra, guerlin, sayrin, zhou}. Nevertheless, in most situations, the system is unavoidable coupled to the environment, that produces detrimental effects on the ideal realization of an experiment. In this way, assuming that the environment is in thermal equilibrium with the system, a cavity mode contains thermal photons on average that have to be removed at the beginning of each experiment. One way to remove the residual thermal photons is sending across the cavity a number of atoms initially prepared in the lower atomic level $|g\rangle$ and tuned in resonance with the cavity mode \cite{nogues, colloquium, guerlin}. However, in order to prepare the cavity field in its vacuum state, a more efficient technique to absorb thermal photons is by the principle of the rapid adiabatic passage (RAP), in which the atom-field frequency is swept \cite{RAP1, RAP2}. These two techniques employ a cooling sequence of atoms to reduce the effective field temperature by energy exchange between the cavity field and the atoms.

In this letter, we present a theorical scheme to cool-down an initial thermal field to the pure vacuum state through an atomic postselection of a sequence of atoms interacting with the field stored in a cavity. Similar works has been proposed for generating Fock states superpositions of a cavity field \cite{oyarce}, creating quantum vibrational states and cooling a nanomechanical oscillator by performing a postselective measurement\cite{montenegro1, montenegro2}. In particular, we consider a dispersive atom-field coupling of each atom in the sequence, and thus there is no energy exchange between the atoms and the field, which is the main difference with previous schemes presented. In the present work, we find the adequate atomic postselection in such a way that the initial thermal field ends up in the vacuum state without energy transfer. Hence, the process is probabilistic due to the conditioned atomic measurements required to obtain the desired state. Additionally, we study the feasibility of our process in an open quantum system under real experimental conditions.
\begin{figure}[h]
\centering \includegraphics[width= \linewidth]{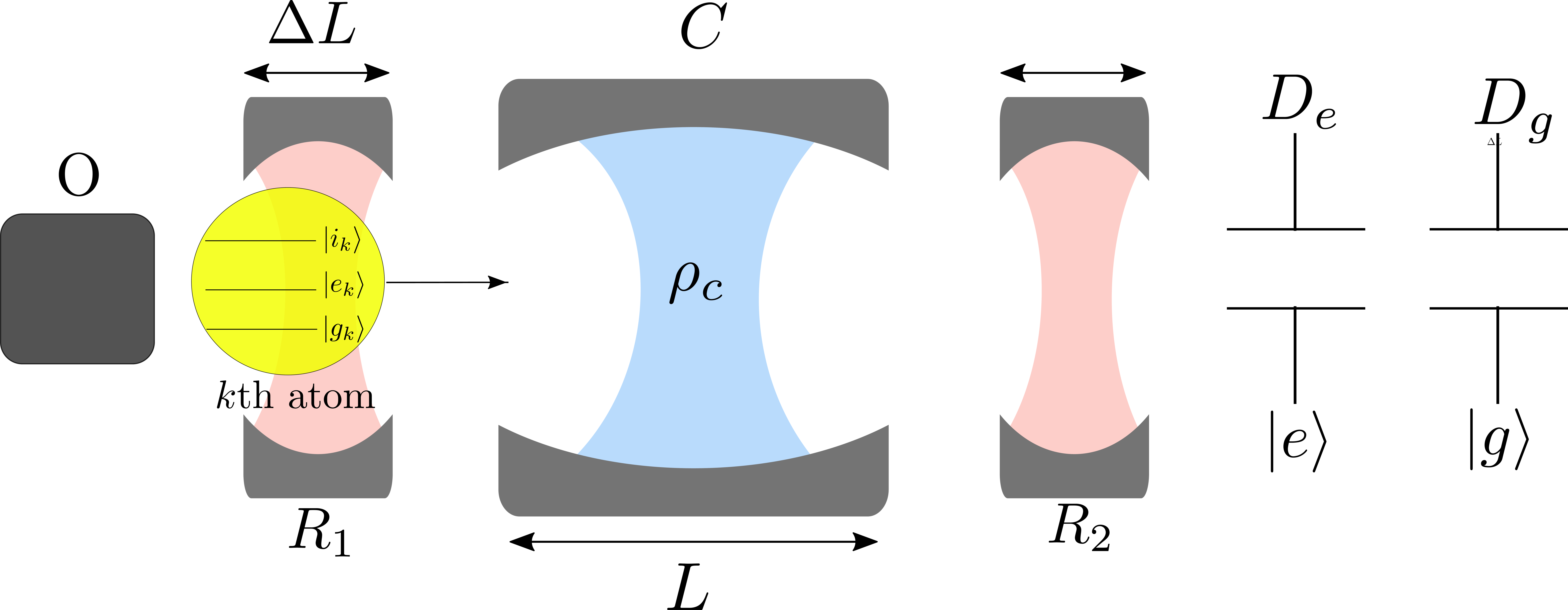}
\caption{(a) Scheme of the cavity QED-setup for Ramsey interferometry. The setup involves a cavity $C$ prepared in an initial state $\rho_c$. In box $O$, the atoms are initially prepared in the state $|e\rangle$ with a given velocity. Then, each atom of the sequence crosses three cavities: $R_1$, $C$ and $R_2$. In each of the $R_1$ and $R_2$ zones, there is a semiclassical interaction atom between each atom and a classical microwave field. This interaction is essential to prepare a superposition of the states $|e\rangle$ and $|g\rangle$ in $R_{1}$, and manipulate the atomic state in $R_{2}$ after the interaction with the field in $C$. At the end of the setup, in the ionization zones $D_e$ and $D_g$, the atomic level is postselected by detecting the atom in the state $|e\rangle$ or $|g\rangle$.}
\label{figure0}
\end{figure}
In the dispersive regime, when the atom-field detuning is large as compared to the coupling, the effective interaction between the atoms and the field produces an energy shift to the atomic state. This energy shift leads to a phase shift on the atomic state which can be measured by a Ramsey interferometer setup as shown in figure \ref{figure0}\cite{QND90}. In the zone $C$, between the classical microwave zones $R_{1}$ and $R_{2}$ used for Ramsey interferometry, we have a superconducting cavity with an initial field given by $\rho_{c}(0) = \sum_{nn'}\rho_{nn'}|n\rangle\langle n'|$ written as an expansion using a Fock basis. The sequence of $N$ three-level atoms is preselected in the state $\rho_{a}(0) = \bigotimes_{k=1}^{N}|e_k\rangle\langle e_k|$ in box $O$ and injected into the setup with a controlled velocity that allows us to assume that there is only one atom flying in the setup at a given time. Then, the initial state of the multipartite system (cavity - atoms) reads as
\begin{equation}
\rho_{ca}(0) = \rho_{c}(0)\otimes\rho_{a}(0).
\end{equation}

The Ramsey interferometer is realized by applying two classical pulses, each one in zones $R_{1}$ and $R_{2}$. The semiclassical interaction between each pulse and the levels $|e_{k}\rangle$ and $|g_{k}\rangle$ of the $k$th atom makes it possible to manipulate the atoms and create linear superpositions of the $|e_{k}\rangle$ and $|g_{k}\rangle$ levels \cite{scully}. Particularly, considering an atomic transition frequency $\omega_{eg}$ resonant with the pulse frequency with a pulse phase of $\phi = \pi/2$ and an interaction time $\Delta\tau_{k}=\Delta L/v_{k}$ which satisfies $\Omega_{R}\Delta\tau_{k}=\pi/2$. The time evolution of the atom, in each zone $R_{1}$ and $R_{2}$, is an unitary transformation $R_{\pi/2}$ given by \cite{exploring}

\begin{equation}
R_{\pi/2} = \frac{1}{\sqrt{2}}\left(\begin{array}{cc} 1 & \rmi\\ \rmi & 1\end{array}\right)
\end{equation}
where $\Delta L$ is the length of the $R_{1}$ and $R_{2}$ zones, $v_{k}$ is the velocity of the atom and $\Omega_{R}$ is the Rabi frequency.

On the other hand, in cavity $C$, we consider the atom as a two-level system that interacts with a quantized mode of a radiation field. In general, this situation is described by the Jaynes-Cummings model
\begin{equation}
H^{(k)} = \frac{\hbar}{2}\omega_{ie}\sigma_{z}^{(k)}+\hbar\omega a^{\dag}a+\hbar g(a\sigma_{+}^{(k)}+a^{\dag}\sigma_{-}^{(k)}).
\end{equation}
Here, the interaction involves only the levels $|i_{k}\rangle$ and $|e_{k}\rangle$ of each atom, and level $|g_{k}\rangle$ does not participate. Consequently, the Pauli operators are $\sigma_{-}^{(k)}=|e_{k}\rangle\langle i_{k}|$, $\sigma_{+}^{(k)}=|i_{k}\rangle\langle e_{k}|$, $\sigma_{z}^{(k)}=|i_{k}\rangle\langle i_{k}|-|e_{k}\rangle\langle e_{k}|$. The single mode of the quantized field is represented by the creation and annihilation operators $a^{\dag}$ and $a$, respectively. As we mentioned above, we are interested in the dispersive regime, when the detuning  between the cavity field frequency $\omega$ and the atomic transition frequency $\omega_{ie}$ is large as compared to the coupling $g$, i.e. $\delta = \omega_{ie}-\omega\gg$ $g\sqrt{n}$. The effective dispersive Hamiltonian in this regime can be written as \cite{scully}
\begin{equation}
H_{eff}^{(k)}= \frac{\hbar g^{2}}{\delta}a^{\dag}a\sigma_{-}^{(k)}\sigma_{+}^{(k)}.
\label{dispersive}
\end{equation}

Equation (\ref{dispersive}) tells us that in both the field and the atom, the number of excitations is conserved, so there is no energy transfer between them. Moreover, during the interaction time, the dispersive Hamiltonian produces a phase shift on  $|e_{k}\rangle$ proportional to the photon number.

Explicitly, the time evolution operator after an interaction time $\tau_{k} = L/v_{k}$ is
\begin{equation}
U_{eff}^{(k)}= \exp\left(-\rmi H_{eff}^{(k)}\tau_{k}/\hbar\right)= \exp\left(-\rmi\varphi_{k} a^{\dag}a\sigma_{-}^{(k)}\sigma_{+}^{(k)}\right),
\end{equation}
where the interaction time $\tau_{k}$ is written in terms of the cavity length $L$ and the velocity $v_{k}$ of the $k$th atom. Also, $\varphi_{k}=g^{2}\tau_{k}/\delta$ is the phase shift of one photon.

Now, following the same general procedure as in \cite{oyarce} for the generation of Fock states superpositions of the field, we can describe the complete evolution of an individual atom that crossed the zones $R_{1}$, $C$ and $R_{2}$ by the evolution operator $U^{(k)}= R_{\pi/2}U_{eff}^{(k)}R_{\pi/2}$. Hence, the total evolution operator of $N$ successive atoms interacting with the cavities after a time $\tau$ is
\begin{equation}
U(\tau) = U^{(N)}...U^{(1)}.
\end{equation}

The procedure is based on a postselection of the atomic levels in a target state $|\psi_{t}\rangle$ over the multipartite state of the whole system ($\rho_{ca}$) after its evolution with the operator $U$. We will see that the main task of our cooling protocol is to determine the different values of $\varphi_k$ combined with a proper postselection.

The postselection of the target state takes place in two ionization zones $D_e$ and $D_g$, where the level of the atoms is detected in $|e\rangle$ or $|g\rangle$. Since the order is not important, we choose a symmetric target state of the form
\begin{equation}
|\psi_t\rangle = \left(\frac{1}{C^{N_e}_N}\right)^{1/2}\sum_{p}|m_{1},...,m_{N}\rangle,\label{postselection}
\end{equation}
with the summation taken over all the possible combinations of $N_{e}$ atoms on the $|e\rangle$ level and $N-N_{e}$ on the $|g\rangle$ level. The normalization factor $C_{N}^{N_{e}}=N!/[N_{e}!(N-N_{e})!]$ is the number of combinations of $N_{e}$ atoms on the $|e\rangle$ level in a set of $N$ atoms.
Therefore, the evolved cavity field state after the atomic postselection is
\begin{equation}
\rho_c(\tau) = \langle\psi_t|U(\tau)\rho_{ca}(0)U^{\dag}(\tau)|\psi_{t}\rangle.
\end{equation}
After some straightforward calculations, the unnormalized field state is
\begin{equation}
\rho_c(\tau)= C_{N}^{N_{e}}\sum_{nn'}\rho_{nn'}\prod_{k=1}^N e^{\frac{\rmi}{2}\varphi_{k}N(n'-n)}c_{n, k}^{N-N_{e}}c_{n', k}^{N-N_{e}}d_{n, k}^{N_{e}}d_{n', k}^{N_{e}}|n\rangle\langle n'|,\label{poststate}
\end{equation}
where the coefficients are $c_{n,k} = \cos(\varphi_{k} n/2)$ and $d_{n,k}=\sin(\varphi_{k} n/2)$. Finally, because our work relies on a postselection process the desired field state is generated with a success probability given by
\begin{equation}\label{postproba}
P_{post} = C_{N}^{N_{e}}\sum_{n}\rho_{nn}\prod_{k=1}^{N}c_{n, k}^{2(N-N_{e})}d_{n, k}^{2N_{e}}.
\end{equation}
 
\begin{figure}[h]
\centering \includegraphics[width= \linewidth]{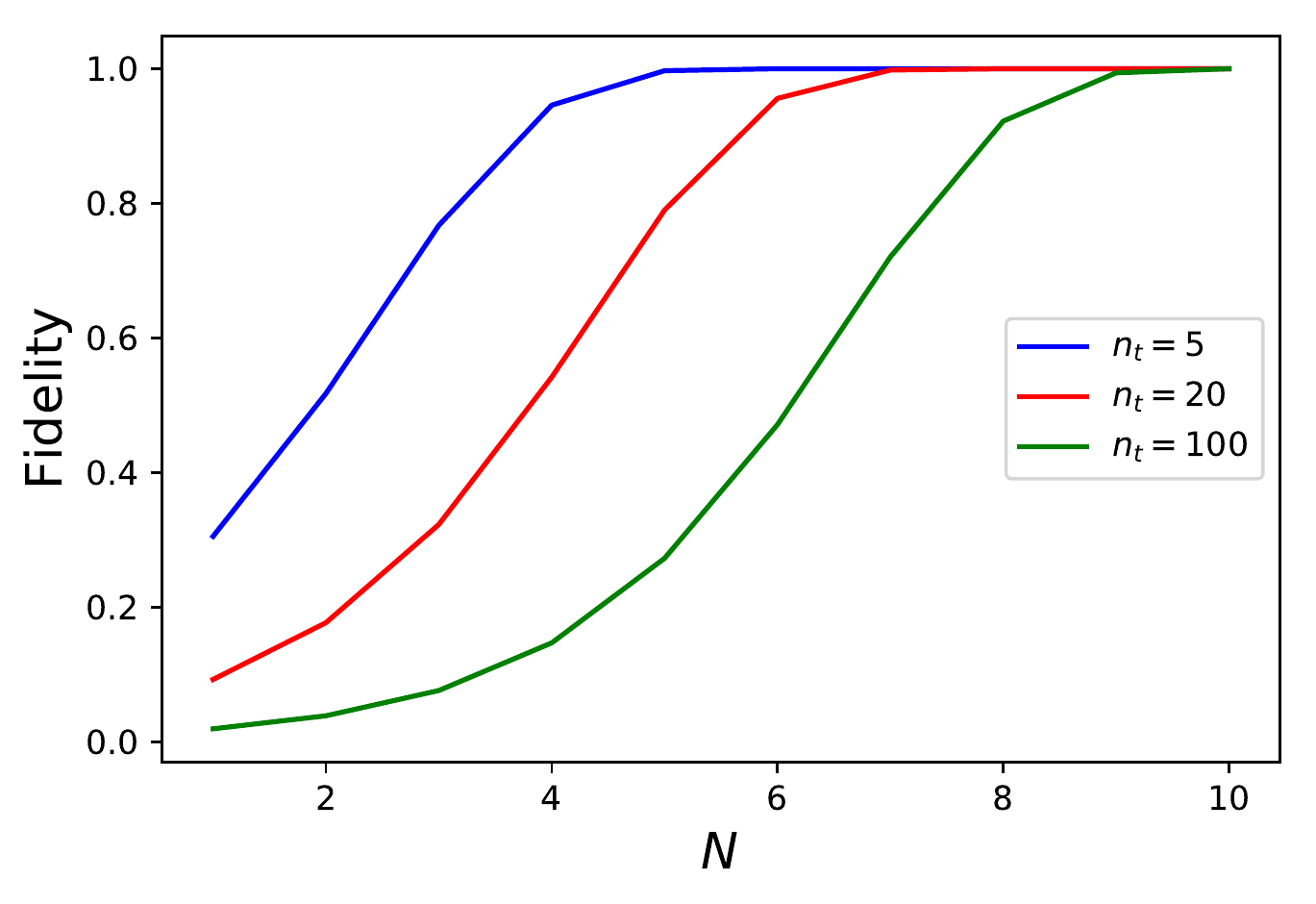}
\caption{Fidelity between the vacuum and the final state of the cavity field given by Eq. (\ref{finalstate}) using a sequence of $N$ atoms interacting with $\varphi_{k}=\pi/2^{k-1}$.}
\label{figure2}
\end{figure}

In the following, we show how to generate the vacuum state $|0\rangle$ of the cavity field by an appropriate atomic postselection starting from a thermal state of the field $\rho_c=\sum_{n}\rho_{nn}|n\rangle\langle n|$, where $\rho_{nn}=n_{t}^n/[(1+n_{t})^{n+1}]$ and $n_{t}$ being the average photon number. As we can see from equation (\ref{poststate}), we should have $N_e=0$ to keep the $|0\rangle$ state. Thus, the normalized state of the field for an initial thermal state after the postselection of $N$ atoms in the $|g\rangle$ level ($N_e=0$) is
\begin{equation}
\rho_f=\frac{\sum_{n}\rho_{nn}\prod_{k=1}^{N}\cos^{2}\left(\varphi_{k} n/2\right)|n\rangle\langle n|}{\sum_{n}\rho_{nn}\prod_{k=1}^{N}\cos^{2}\left(\varphi_{k} n/2\right)},
\label{finalstate}
\end{equation}
with postselection probability
\begin{equation}
P_{post}=\sum_{n=0}^{\infty}\rho_{nn}\prod_{k=1}^{N}\cos^{2}\left(\frac{\varphi_{k}n}{2}\right).
\label{postprob}
\end{equation}
As shown in equation (\ref{finalstate}), the parameters $\varphi_{k}$ have to be adequate to ensure that the oscillatory function $\prod_{k=1}^{N}\cos^2\left(\varphi_{k}n/2\right)$ multiplying the projectors $|n\rangle\langle n|$ is close to zero for all the photon numbers except $n=0$. We propose a sequence of atoms where the $k$th atom crosses with $\varphi_{k}=\pi/2^{k-1}$ in order to eliminate the photon numbers $n = (2m-1)2^{k}$. Once the process has finished the postselection probability is $P_{post} \to \rho_{00}=1/(1+n_t)$. Figure \ref{figure2} shows the fidelity between the vacuum state and the final state of the cavity field after the postselection of a sequence of $N$ atoms in $|g\rangle$. For an initial state with a mean photon number $n_{t}=100$, the cooling process converges after the detection of a sequence of about 10 atoms.

\begin{figure}
\centering \includegraphics[width=\linewidth]{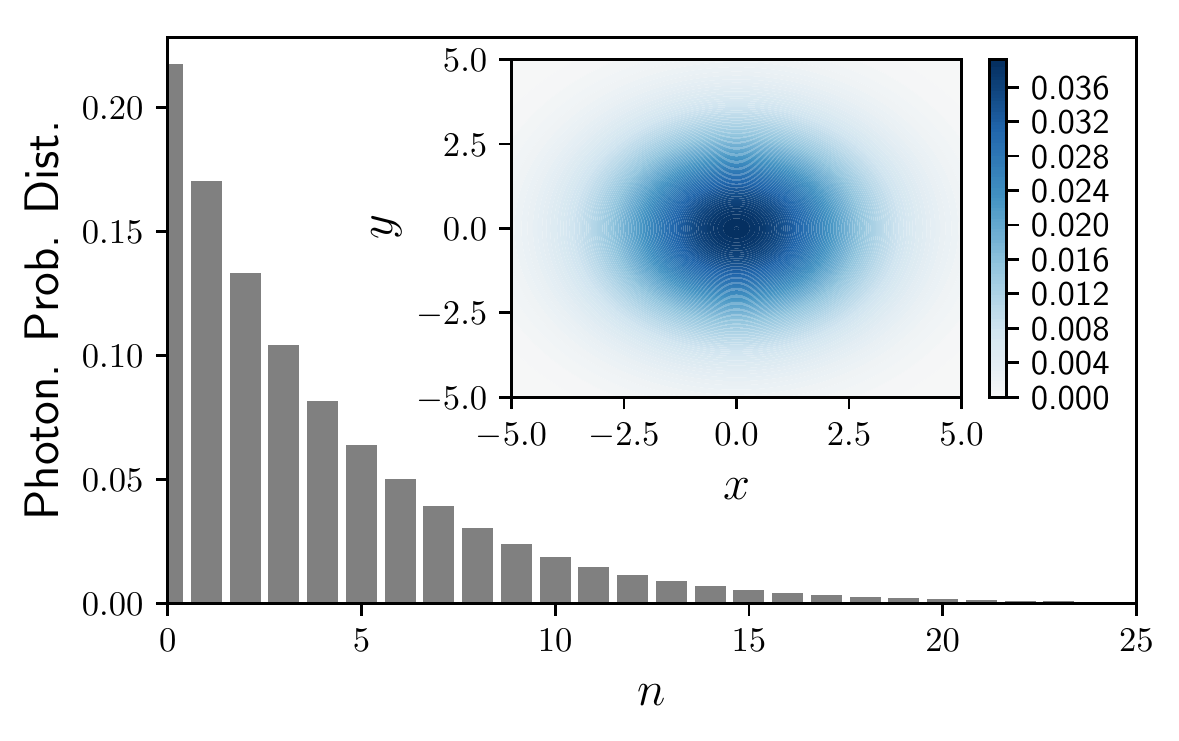}
\centering \includegraphics[width=\linewidth]{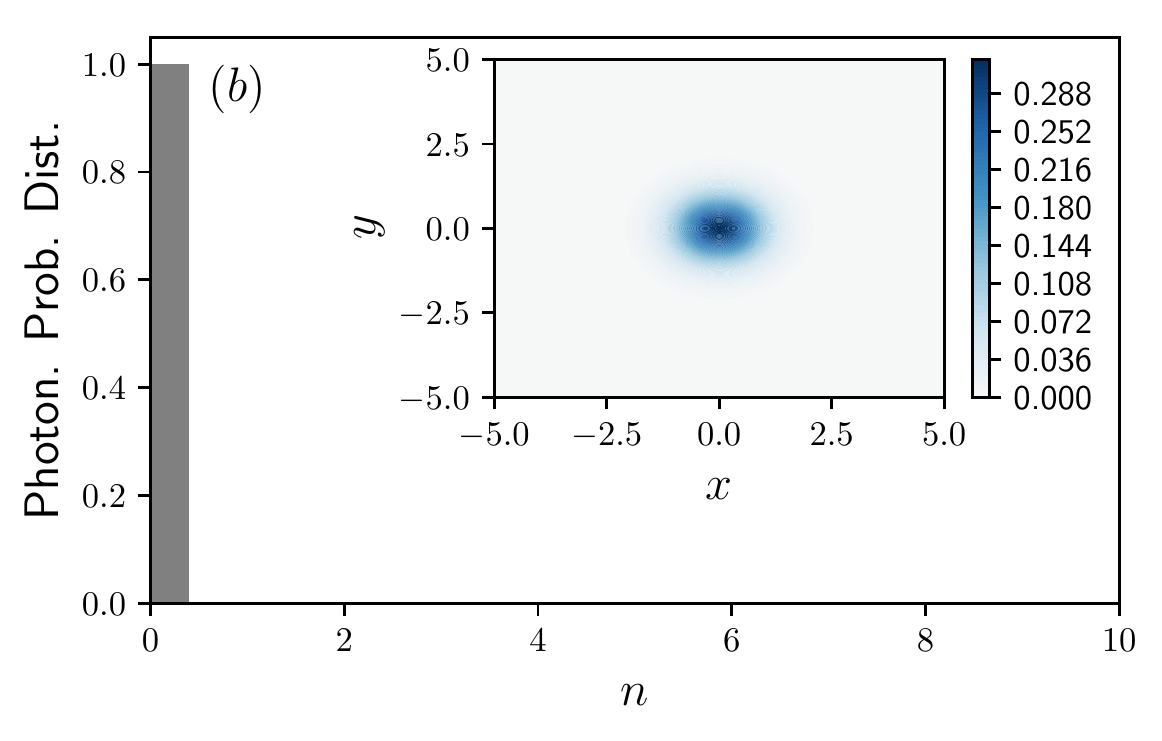}
\caption{Cooling process of the thermal field. In (a) we show the initial initial thermal state ($n_{t}=3.6$), whereas in the bottom panel (b), we show the final state of the field after the postselection of a sequence of 5 atoms in $|g\rangle$. In both figures the photon number distribution and Wigner function are shown. The vacuum photon state is generated with a probability of $P_{post}\approx21.7\%$.}
\label{figure3}
\end{figure}

In figure \ref{figure3}, we illustrate the convergence of the sequence with $\varphi_{k}=\pi/2^{k-1}$, considering that all the atoms are postselected in the state $|g\rangle$. In the top panel (a), we plotted an initial thermal state with $n_t=3.6$, whereas in the bottom panel (b), we show the final field state after the postselection of 5 atoms in the state $|g\rangle$. As seen, the final state is the vacuum photon state $|0\rangle$, where the Wigner function $W^{[\rho_{t}]}(\alpha) = \frac{2}{\pi}\frac{1}{2n_{t}+1}e^{-2|\alpha|^{2}/(2n_{t}+1)}$ becomes sharper than the initial thermal state.

In a realistic scenario, the quantum system is coupled to the environment and suffers from decoherence effects.
Since we are cooling photons, we are fighting against the thermalization effect of the reservoir.
To simulate this scenario, we consider that the cavity field is initially in thermal equilibrium with an average photon number $n_t$, and that the atom-field system evolves with the following master equation:
\begin{equation}
\frac{d\rho_{S}}{dt}=\frac{1}{i\hbar}[H_{eff},\rho_{S}]+\sum_{i}\left[L_{i}\rho_{S}L_{i}^{\dag}-\frac{1}{2}(L_{i}^{\dag}L_{i}\rho_{S}+\rho_{S}L_{i}^{\dag}L_{i})\right],\label{mastereq}
\end{equation}
where $H_{eff}$ is the dispersive coupling Hamiltonian of equation (\ref{dispersive}) for each atom in the sequence and the $L_{i}$ Lindbland operators are $\sqrt{\Gamma(1+n_t)}\sigma_{-}$, $\sqrt{\Gamma n_t}\sigma_{+}$, $\sqrt{\kappa(1+n_t)}a$ and $\sqrt{\kappa n_t}a^{\dag}$. 

Typically, in microwave experiments with circular Rydberg atoms, the relaxation of these atoms is negligible, when compared to the cavity damping time $T_c$.
However, we did consider the lifetime reduction by a factor $(1+n_t)$, as well as the field losses, but neglected the losses during the transit time through the Ramsey zones.
The master equation for the period without atoms ($t\sim82$ $\mu$s), is given by:
\begin{equation}
\frac{d\rho_{c}}{dt}=\sum_{i}\left[L_{i}\rho_{S}L_{i}^{\dag}-\frac{1}{2}(L_{i}^{\dag}L_{i}\rho_{S}+\rho_{S}L_{i}^{\dag}L_{i})\right],
\end{equation}
where the Lindbland operators are $\sqrt{\kappa(1+n_t)}a$ and $\sqrt{\kappa n_t}a^{\dag}$. In our calculations, we used a cavity damping time $T_c = 1/\kappa=130$ ms \cite{kuhr}. Also, the atomic lifetime is $T_a=1/\Gamma=30$ ms for circular Rydberg atoms of rubidium with principal quantum numbers 51 or 50 \cite{exploring}. We consider the cavity tuned at a frequency $\omega/2\pi=51.1$ GHz and an atom-cavity detuning $\delta/2\pi=245$ kHz. The vacuum Rabi frequency is $g/2\pi=49$ kHz. All of these parameters are consistent with real experimental realizations \cite{sayrin}.

\begin{figure}
\centering \includegraphics[width=\linewidth]{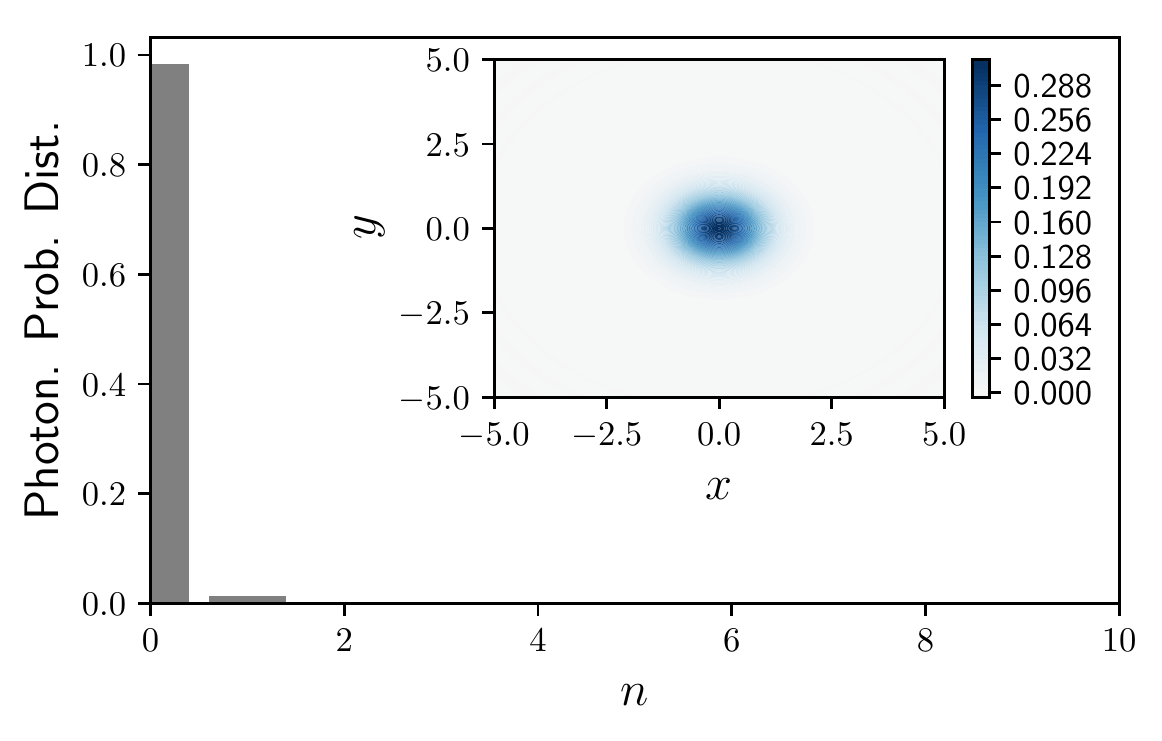}
\caption{Cooling process of figure \ref{figure3} considering thermal effects of the environment at a temperature $T=10$ K (mean photon number $n_{t}=3.6$). The final state is close to a thermal state with a mean photon number $n_{t}=0.017$, corresponding to a temperature $T=0.6$ K.}
\label{figure4}
\end{figure}

The temperature of a thermal state can be determined by the relation $n_{t}=1/[\exp{(\hbar\omega/k_{b}T)}-1]$. Hence, for an initial thermal state with $n_t=3.6$ and the above given frequency, the corresponding bath and photon temperature is $T=10$ K (figure \ref{figure3}a).
After the cooling process and considering the effect of the reservoir, the fidelity between the final state and vacuum goes down to 98.3$\%$ (figure \ref{figure4}). In this case, the final state has a 99.7$\%$ fidelity with respect to a thermal state with $n_{t}=0.017$ corresponding to a temperature of $T=0.6$ K. This result shows that the cooling process is robust even in the presence of decoherence, considered by the master equation in equation (\ref{mastereq}).

As we mentioned above, this process can be done in a typical CQED system using circular Rydberg atoms. These are the Rydberg levels with the highest angular momentum and have a very long lifetime, on the order of tens of miliseconds. We can also neglect the decoherence of the atoms as they fly through the setup due to the short interaction time ($\sim 0.4$ miliseconds). In this type of experiment, all the parameters of the atomic samples are under control such as velocity, preparation time and interaction time. Therefore, the couplings $\varphi_{k}$ proposed in our atomic sequence to cool down the thermal photons in the cavity by atomic postselection can be achieved by controlling the velocity of each atom by laser techniques. However, our scheme requires that each atomic sample contains deterministically only one atom, and this is challenging in real experiments. Most of the experiments prepare the atomic samples by weak laser excitation resulting in a Poissonian distribution for the number of atoms in the sample. Despite of this, we assume in this work a deterministic preparation of single atoms considering some schemes proposed to achieve the single-atom preparation of Rydberg atoms using the called $\textit{dipole blockade}$ effect \cite{dipoleblockade1, dipoleblockade2}.

In summary, in this work we suggest a protocol to cool-down a thermal field to its vacuum state using a typical cavity QED-setup. Here, a sequence of atoms is sent interacting (one at a time) dispersively with the cavity field. After the interaction the atoms are postselected in its ground state, so our protocol is probabilistic. This means that if we obtain a different result than expected in an atomic detection, we have to fully reinitialize the scheme to achieve our goal, with a certain success probability. In order to accomplish our task with a minimum number of atoms, we propose a sequence interacting with $\varphi_{k}=\pi/2^{k-1}$ which rapidly eliminate the nonzero photon components. The reduction of the number of atoms needed in the process is important when the relevant system is coupled to a thermal reservoir with $T\neq0$, since the whole process takes less time. We model this situation using the general master equation in equation (\ref{mastereq}), where we have considered atomic and field losses by taking real experimental parameters. Finally, we conclude that even in the presence of decoherence our protocol can be done with high fidelity.

\section*{Acknowledgments}
MO acknowledges the financial support of the project Fondecyt (1180175).

\end{document}